\documentclass[twocolumn,preprintnumbers,amsmath,amssymb]{revtex4-1}

\usepackage{graphicx}
\usepackage{dcolumn}
\usepackage{bm,epsfig}
\usepackage{hyperref}
\usepackage{epstopdf}
\usepackage{amsfonts}
\usepackage{color}
\setcitestyle{super}
\usepackage[utf8]{inputenc}
\usepackage{fancyhdr}
\begin{document}

\title{Colossal magnetoresistance in a nonsymmorphic antiferromagnetic insulator}

\author{P. F. S. Rosa$^{1}$, Yuanfeng Xu$^{2}$, M. C. Rahn$^{1}$, J. C. Souza$^{3}$, S. K. Kushwaha$^{1,4}$, L. S. I. Veiga$^{6}$, A. Bombardi$^{7}$, 
S. M. Thomas$^{1}$, M. Janoschek$^{5}$, E. D. Bauer$^{1}$,  M. K. Chan$^{4}$, Zhijun Wang$^{9}$, J. D. Thompson$^{1}$, N. Harrison$^{4}$, P. G. Pagliuso$^{3}$, B. A. Bernevig$^{8}$,
F. Ronning$^{1}$}
\affiliation{
$^{1}$ Quantum Group, Los Alamos National Laboratory, Los Alamos, NM 87545, USA.\\
$^{2}$ Max Planck Institute of Microstructure Physics, 06120 Halle, Germany.\\
$^{3}$ Instituto de F\'{\i}sica ``Gleb Wataghin'', UNICAMP, 13083-859, Campinas, SP, Brazil.\\
$^{4}$ National High Magnetic Field Laboratory, Los Alamos National Laboratory, MS E536, Los Alamos, NM 87545, USA.\\
$^{5}$  Laboratory for Neutron and Muon Instrumentation, Paul Scherrer Institut, Villigen, Switzerland.\\
$^{6}$ Department of Physics and Astronomy, University College London, Gower Street, London WC1E 6BT, UK.\\
$^{7}$ Diamond Light Source, Harwell Science and Innovation Campus, Didcot, OX11 0DE, UK.\\
$^{8}$ Department of Physics, Princeton University, Princeton, NJ 08544, USA.\\
$^{9}$ Beijing National Laboratory for Condensed Matter Physics, and Institute of Physics, Chinese Academy of Sciences, Beijing 100190, China 
\& University of Chinese Academy of Sciences, Beijing 100049, China.}
\date{\today}

\begin{abstract}
Here we investigate antiferromagnetic Eu$_{5}$In$_{2}$Sb$_{6}$, a nonsymmorphic Zintl phase.
Our electrical transport data show that Eu$_{5}$In$_{2}$Sb$_{6}$ is remarkably insulating and
exhibits an exceptionally large negative magnetoresistance, 
which is consistent with the presence of magnetic polarons. 
From {\it ab initio} calculations, the paramagnetic state of Eu$_{5}$In$_{2}$Sb$_{6}$ 
is a topologically nontrivial semimetal within the generalized gradient approximation (GGA), 
whereas an insulating state with trivial topological indices is obtained using a modified
Becke-Johnson potential. Notably, GGA+U calculations suggest that
the antiferromagnetic phase of Eu$_{5}$In$_{2}$Sb$_{6}$ may host an axion insulating state.
Our results provide important feedback for theories of 
topological classification and highlight the 
potential of realizing clean magnetic narrow-gap semiconductors in 
Zintl materials.
\end{abstract}

\maketitle

\section{Introduction}
\vspace{-0.5cm}
Narrow-gap semiconductors exhibit a breadth of striking functionalities
ranging from thermoelectricity to dark matter detection \cite{Tomczak2018,Hochberg2018}. 
More recently, 
the concept of topological insulating phases in bulk materials has renewed the interest in this class 
of materials \cite{Hasan2010, Moore2010,Hasan2011}.
Independent of the target application, a primary goal from the experimental point of view
is the synthesis of genuine insulators free of 
self doping. Materials design is usually guided by simple electron count (e.g. tetradymite Bi$_{2}$Te$_{3}$ \cite{Cava2013}),
correlated gaps (e.g.~Kondo insulators SmB$_{6}$\cite{Eo2019a} and YbB$_{12}$\cite{Sato2019}) 
or the Zintl concept (e.g.~Sr$_{2}$Pb \cite{Sun2011} and BaCaPb \cite{Wang2018}). 
Zintl phases are valence precise intermetallic phases
formed by cations (alkaline, alkaline-earth and rare-earth elements)
and covalently bonded (poly)anionic structures containing post-transition metals. 
The electron transfer between these two entities gives rise to
an insulating state, whereas the inclusion of rare-earth elements allows for magnetism, which
breaks time-reversal symmetry and may promote new quantum ground states \cite{Chang2020,Yu2010,Zhang2019b}.

The myriad of crystal structures within the Zintl concept provides a promising avenue to search for 
clean semiconductors. Here we experimentally investigate Zintl Eu$_{5}$In$_{2}$Sb$_{6}$ 
in single crystalline form. 
Low-carrier density magnetic materials containing Europium are prone to exhibiting colossal
magnetoresistance (CMR) \cite{Shapira1972,Chan1997,Chan1998,Sullow2000,Devlin2018}.  
The strong exchange coupling between the spin of the carriers and the spins of the Eu$^{2+}$ background 
causes free carriers at low densities to self-trap in ferromagnetic clusters around the Eu sites, 
which gives rise to a quasiparticle called magnetic polaron \cite{Kasuya1968}. 
This quasiparticle has been identified in several Zintl materials ranging from simple cubic 
EuB$_{6}$ \cite{Sullow2000,Pohlit2018}
to monoclinic Eu$_{11}$Zn$_{4}$Sn$_{2}$As$_{12}$ \cite{Devlin2018}. 
Most CMR compounds have a ferromagnetic
  ground state, including doped magnanites $RE_{1-x}A_{x}$MnO$_{3}$ ($RE=$ rare-earth, $A=$ divalent cation)
  in which CMR was first observed \cite{Jin1994,Salamon2001}. 
  EuTe and Eu$_{14}$MnBi$_{11}$, however, revealed the possibility of realizing CMR in 
 antiferromagnets \cite{Shapira1972,Chan1998,Sakurai2013}, which also brings promise for applications due to their small 
 stray fields \cite{Baltz2018}.

Additionally, nonsymmorphic symmetries are expected to be
 particularly powerful in creating protected band
crossings and surface states, which provide an additional organizing principle 
within the Zintl concept \cite{Young2015,Parameswaran2013}.
For instance, Wieder \textit{et al} predicted that Zintl 
Ba$_{5}$In$_{2}$Sb$_{6}$, the non-$f$ analog of Eu$_{5}$In$_{2}$Sb$_{6}$, 
hosts fourfold Dirac
fermions at $\bar{M}$ connected to an
hourglass fermion along $\bar{\Gamma} \bar{X}$ \cite{Wieder2018}. Recent attempts to theoretically 
catalogue all known 
uncorrelated materials indicate that Ba$_{5}$In$_{2}$Sb$_{6}$ may be classified as a 
topological insulator \cite{Bradlyn2017, Vergniory2019}
or trivial insulator \cite{Zhang2019,Tang2019}.
This discrepancy begs for an experimental investigation.


 Eu$_{5}$In$_{2}$Sb$_{6}$, just like its Ba analog, crystallizes in
 space group $Pbam$. As expected from the $4f$ localized moments in multiple sites,
 Eu$_{5}$In$_{2}$Sb$_{6}$ orders antiferromagnetically at $T_{\mathrm{N1}}$=14 K in a complex magnetic structure.
 Remarkably, colossal magnetoresistance sets in at $15T_{\mathrm{N1}}$ and is accompanied by an anomalous 
 Hall component. Our data collectively point to the presence of magnetic polarons. To shed light on the topology of the
  band structure of Eu$_{5}$In$_{2}$Sb$_{6}$, 
 we have performed first-principles calculations using different
functionals and magnetic phases. 
 Though an insulating state with trivial topological
  indices is obtained using modified Becke-Johnson (mBJ) functional in the paramagnetic state, 
  topological non-trivial states with strong indices emerge in the 
  generalized gradient approximation (GGA)+U calculations 
  within putative antiferromagnetic states.
  \vspace{-0.5cm}
\section{Results}
\vspace{-0.5cm}
\textbf{Magnetic susceptibility measurements}

We first discuss the thermodynamic properties of Eu$_{5}$In$_{2}$Sb$_{6}$ single crystals. Figure~1a
highlights the complex anisotropy in the low-temperature magnetic susceptibility of Eu$_{5}$In$_{2}$Sb$_{6}$. Two 
magnetic transitions can be identified at $T_{\mathrm{N1}}$=$14$~K and $T_{\mathrm{N2}}$=$7$~K, in agreement with 
previous measurements on polycrystalline samples \cite{Subbarao2016}. 
One can also infer that the $c$-axis is the magnetization hard-axis and that the moments lie in the $ab$-plane.
No hysteresis is observed between zero-field-cooled and field-cooled measurements at 0.1~T, 
which rules out hard ferromagnetic order or spin-glass behavior; however, a small in-plane ferromagnetic component 
($0.06~\mu_{\mathrm{B}}$) is observed at 
very low fields ($B \leq 0.1$~T), indicative of a complex magnetic structure with canted moments (see Supplementary Figure 1).

The inset of Fig.~1a shows the product of magnetic susceptibility and temperature as a function of temperature.
At high temperatures ($T>225$~K), a Curie-Weiss (CW) fit yields a ferromagnetic (FM) Weiss temperature of $\theta$=$30$~K 
despite the antiferromagnetic (AFM) order at 
low temperatures, which further corroborates the presence of a complex magnetic configuration with multiple exchange 
interactions. The inverse of the magnetic susceptibility is shown in Supplementary Figure 3. The CW fit also yields an effective moment of 
$8$~$\mu_{\mathrm{B}}$Eu$^{-1}$, in good agreement with the Hund's rule
moment of $7.94$~$\mu_{\mathrm{B}}$Eu$^{-1}$ for Eu$^{2+}$. In fact, our x-ray absorption spectra at the Eu $L$ edges\cite{Ruck2011}
confirm that all three Eu sites are divalent (see Supplementary Figure 5). 
Previous x-ray absorption studies observed
a finite Eu$^{3+}$ component, which could be due to an impurity phase present in polycrystalline 
samples \cite{Subbarao2016}. The fully divalent character of europium 
in Eu$_{5}$In$_{2}$Sb$_{6}$ has been recently confirmed by Mossbauer measurements \cite{Radzieowski2020}.

Notably, our magnetic susceptibility data deviate from the CW fit at temperatures well above the ordering temperature (inset of Fig.~1a).
In purely divalent compounds such as Eu$_{5}$In$_{2}$Sb$_{6}$, 
Eu$^{2+}$ is a localized $S$-only ion ($J=S=7/2$), which implies crystal-field and Kondo effects to be negligible to first order.
As a result, the deviation from a CW fit indicates the presence of short-range magnetic interactions as 
 observed previously in the 
 manganites $RE_{1-x}A_{x}$MnO$_{3}$ ($RE=$ rare-earth, $A=$ divalent cation). 
 Based on small-angle neutron scattering measurements, 
 this deviation was argued to be due to the formation of magnetic polarons \cite{Teresa1997}. 
As temperature decreases, magnetic polarons are expected to grow in size and eventually overlap 
 when $n\xi^{3} \approx 1$, where $n$ is the carrier density
 and $\xi$ is the magnetic correlation length \cite{Majumdar1998}.
 The inset of Fig.~1a shows a sharp decrease in $\chi(T)T$ at $T^{*} \sim 40$~K, which reflects the
 onset of strong antiferromagnetic 
correlations between polarons.

Figure 1b shows the low-temperature anisotropic magnetization of
Eu$_{5}$In$_{2}$Sb$_{6}$. The hard $c$-axis 
magnetization increases linearly with field, whereas 
a field-induced transition is observed within the basal plane
 before saturation is reached at about $10$~T (inset of Fig.~1b). Figure 1c shows the temperature 
dependence of the specific heat, $C$, at zero field. In agreement with magnetic susceptibility data, $C/T$ 
exhibits two phase transitions at $T_{\mathrm{N1}}$ and $T_{\mathrm{N2}}$ as well as a magnon contribution below $T_{N2}$, typical of 
Eu$^{2+}$ compounds. The entropy recovered at $T_{\mathrm{N1}}$ 
is about $90$\% of $R$ln$8$ (not shown), the expected entropy from the 
Eu$^{2+}$ ($J=7/2$) ground state. The 
 extrapolation of the zero-field $C/T$ to $T=0$ gives a Sommerfeld coefficient of zero within the experimental error, indicating that 
 Eu$_{5}$In$_{2}$Sb$_{6}$ is an insulator with very small amounts of impurities. A Schottky-like anomaly at about 35~K indicates
 the presence of short-range correlations, in agreement with magnetic susceptibility data at $T^{*}$ .
 The inset of Fig.~1c displays the field 
dependence of the low-temperature transitions when field is applied along the $b$-axis. The transitions are mostly suppressed by $9$~T, 
in agreement with the saturation in magnetization.
 
    \begin{figure}[!ht]
 \begin{center}
 \includegraphics[width=1\columnwidth]{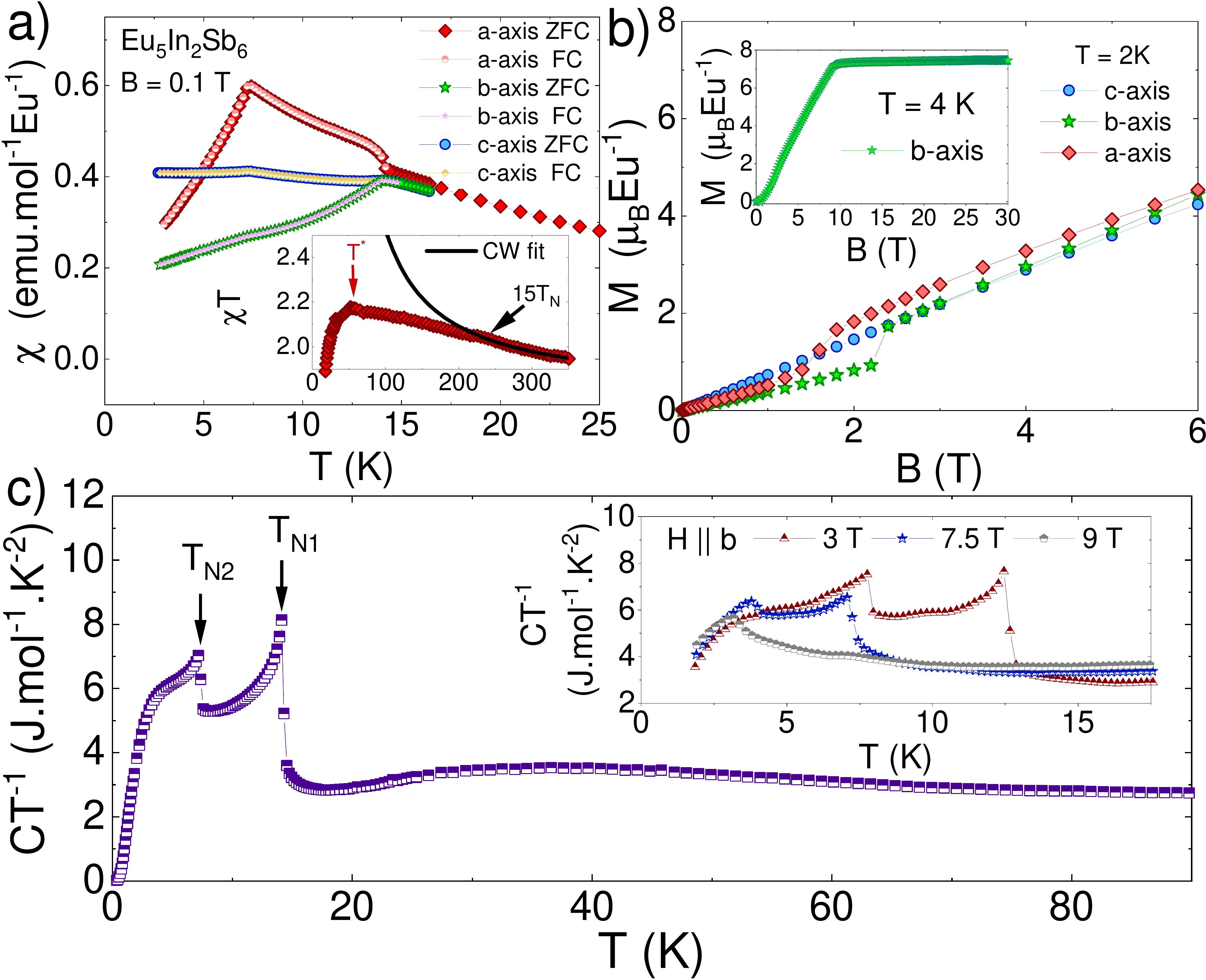}
 \vspace{-0.5cm}
 \end{center}
 \caption{\textbf{Thermodynamic properties of Eu$_{5}$In$_{2}$Sb$_{6}$ crystals.} a) Magnetic susceptibility, $\chi(T)$, 
 in both zero-field-cooled (ZFC) and field-cooled (FC) sweeps. Inset shows $\chi T$. Black solid line shows the high-temperature CW fit. 
 b) Magnetization $vs$ applied field at 2~K. Inset shows high-field magnetization data at 4~K. 
 c) Zero-field specific heat as a function of temperature. Inset shows $C/T$ at different applied fields.}
 \label{fig:Fig1}
 \end{figure}

 \begin{figure*}[!ht]
 \begin{center}
 \includegraphics[width=2\columnwidth]{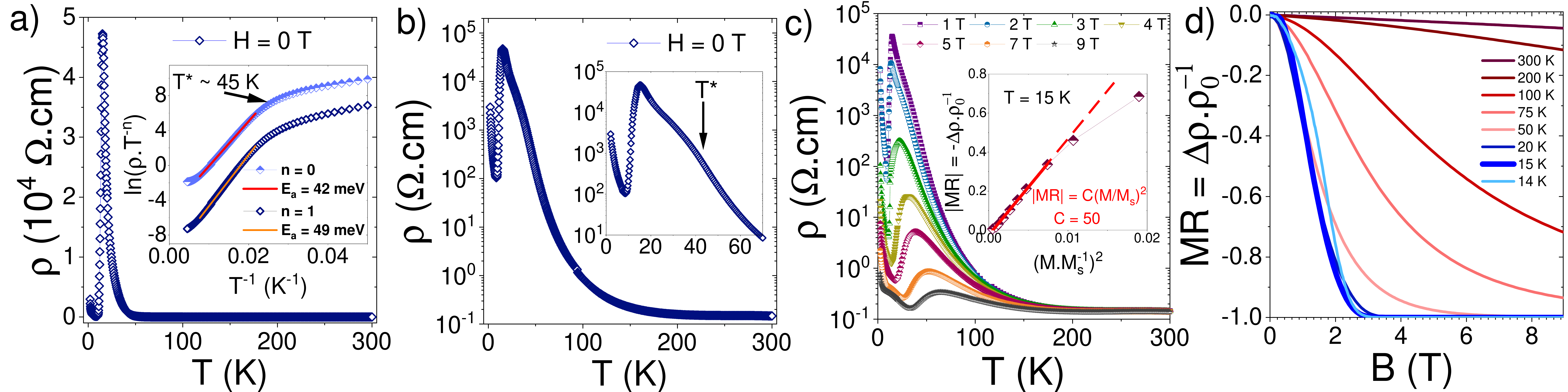}
 \vspace{-0.25cm}
 \end{center}
 \caption{\textbf{Electrical resistivity measurements on Eu$_{5}$In$_{2}$Sb$_{6}$ single crystals.}  a) Temperature dependent electrical resistivity, $\rho(T)$, at zero field. Electrical current was applied along the $c$-axis. 
 Inset shows an activated log $\rho$ $vs$ $1/T$ plot. b) $\rho(T)$ in a log plot. Inset shows a zoom in the  low-temperature region. c) $\rho(T)$ at various magnetic fields applied long the $b$-axis. 
 Inset shows MR $vs$ reduced magnetization ($M/M_{s}$) squared. d) MR $vs$ applied field at various temperatures.} 
 \label{fig:Fig2}
 \end{figure*}

\textbf{Electrical transport measurements}

We now turn our attention to electrical transport data. Figure~2a shows the temperature dependent electrical 
resistivity, $\rho(T)$, of Eu$_{5}$In$_{2}$Sb$_{6}$ measured with current along the $c$-axis.
Remarkably, $\rho(T)$ rises by almost six orders of magnitude in the paramagnetic state, in agreement 
with the clean insulating response observed in $C/T$ but in stark contrast to $\rho(T)$ measurements in 
polycrystals \cite{Cordier1988}. 
Below $T_{\mathrm{N1}}$, $\rho(T)$
decreases by three orders of magnitude, pointing to the overlap of magnetic polarons within the 
antiferromagnetic state. Finally, at lower temperatures $\rho(T)$ rises again,
and a small kink is observed 
at $T_{\mathrm{N2}}$.

The high-temperature electrical resistivity can be fit to an activated behavior given 
by $\rho_{0}\tilde{T}^{n}\mathrm{exp}(E_{\mathrm{a}}/k_{\mathrm{B}}T)$ (inset of Fig.~2a),
where $\tilde{T}$ is the reduced temperature. 
For $n=0$, the Arrhenius plot
yields a narrow gap of $40$~meV whereas a slightly larger energy
is extracted when $n=1$ 
for adiabatic small-polaron hopping conduction \cite{Emin1969}. 
From these data alone, it is not possible
to differentiate between the two mechanisms. 
Nevertheless, the activated behavior breaks down
at about $T^{*}$$\sim$$40$~K, indicating that another mechanism is present. 
This energy scale is more pronounced in 
a log plot shown in Fig.~2b. 

The evolution of the colossal magnetoresistance in Eu$_{5}$In$_{2}$Sb$_{6}$ is summarized in Figure~2d.
Though the negative magnetoresistance is small at room temperature, it rapidly increases below about $15T_{\mathrm{N1}}$. 
At liquid nitrogen 
temperatures ($T \sim 75$~K), for instance, the MR reaches $-50$\% at only 3~T and $-94$\% at 9~T.
Ultimately, the MR peaks at $-99.999$\% at 9~T and $15$~K. This is, to our knowledge, the largest 
CMR observed in a stoichiometric antiferromagnetic compound. 

Hall measurements provide valuable information on the type of carriers and the scattering mechanisms 
in a material. 
Figure~3 shows the Hall resistivity, $R_{\mathrm{H}}\equiv  \rho_{xz}$, for fields applied along the 
$b$-axis of Eu$_{5}$In$_{2}$Sb$_{6}$. At room temperature, 
$R_{\mathrm{H}}$ is linear, as expected from a nonmagnetic single-band material (inset of Fig.~3a). 
The positive slope, $R_{\mathrm{0}}$, 
implies positive (hole) carriers and a carrier density of
 $n_{\mathrm{h}} = 1/R_{\mathrm{0}}e=10^{17}/$cm$^{3}$, typical of narrow-gap semiconductors.

 \begin{figure}[!h]
 \begin{center}
 \includegraphics[width=0.9\columnwidth,keepaspectratio]{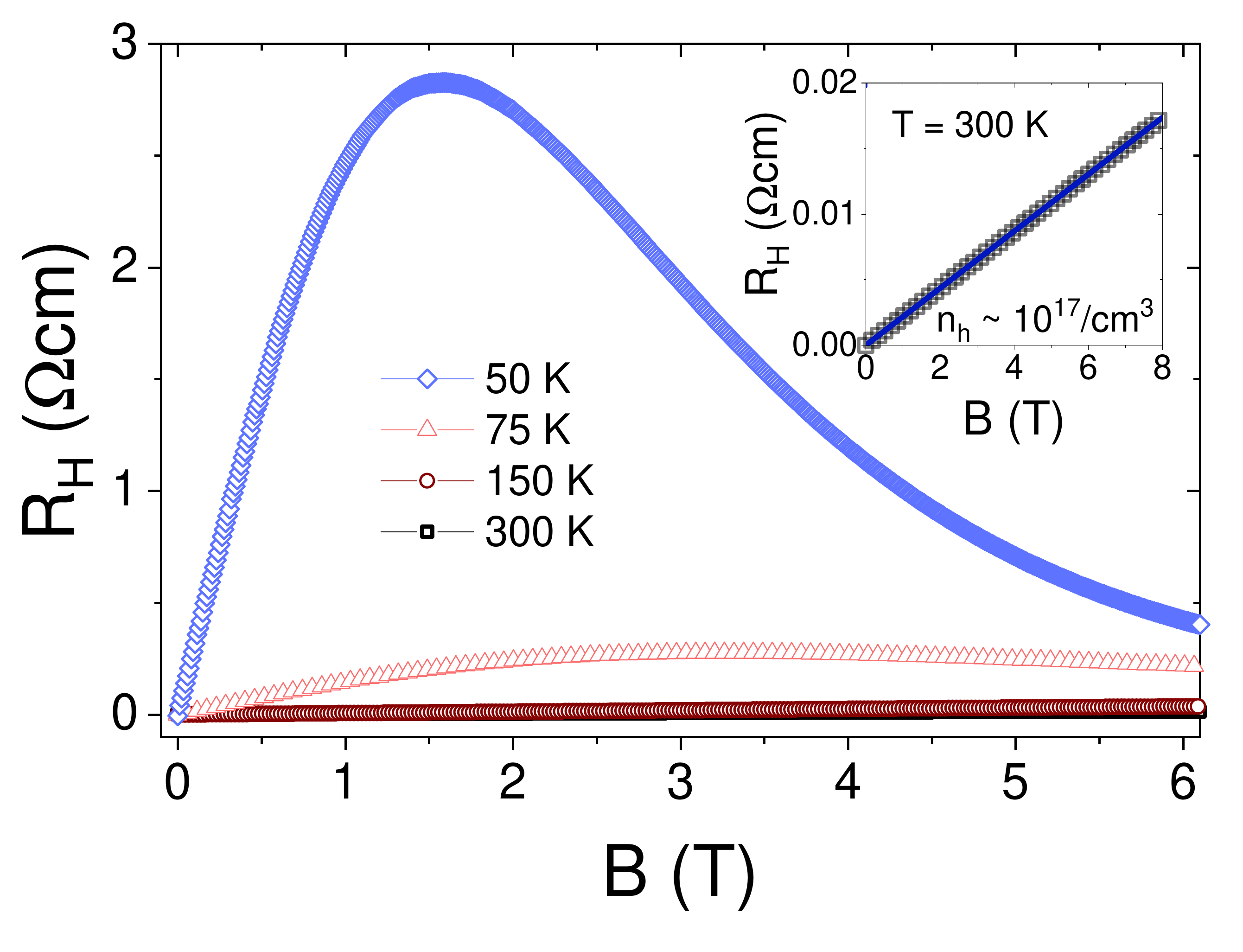}
 \end{center}
 \vspace{-0.75cm}
 \caption{\textbf{Hall effect of Eu$_{5}$In$_{2}$Sb$_{6}$ single crystals.} Hall resistivity $vs$ applied field at various temperatures. Current is applied along the $c$-axis and fields
 are along the $b$-axis. Inset shows the linear Hall response at $300$~K.}
 \vspace{-0.2cm}
 \label{fig:Fig3}
 \end{figure}

As the temperature
  is lowered, however, a 
 nonlinear $R_{\mathrm{H}}$ component sets in at about $15T_{\mathrm{N1}}$, the same temperature 
at which CMR emerges. As the band structure of this band insulator is not expected to change dramatically in this 
temperature range, our result may indicate that the formation of magnetic polarons is responsible for the anomalous Hall 
effect (AHE). We note, however, that the presence of multiple carriers 
cannot be ruled out at this time. Though the ferromagnetic nature of the
magnetic polaron cluster is a natural explanation for the
anomalous contribution, a quantitative analysis
of the various intrinsic and extrinsic contributions to the AHE
will require determining the anisotropic conductivity tensor
 using micro-fabricated devices, including the region below 50~K.
 
  \begin{figure}[!ht]
   \begin{center}
   \includegraphics[width=1.0\columnwidth,keepaspectratio]{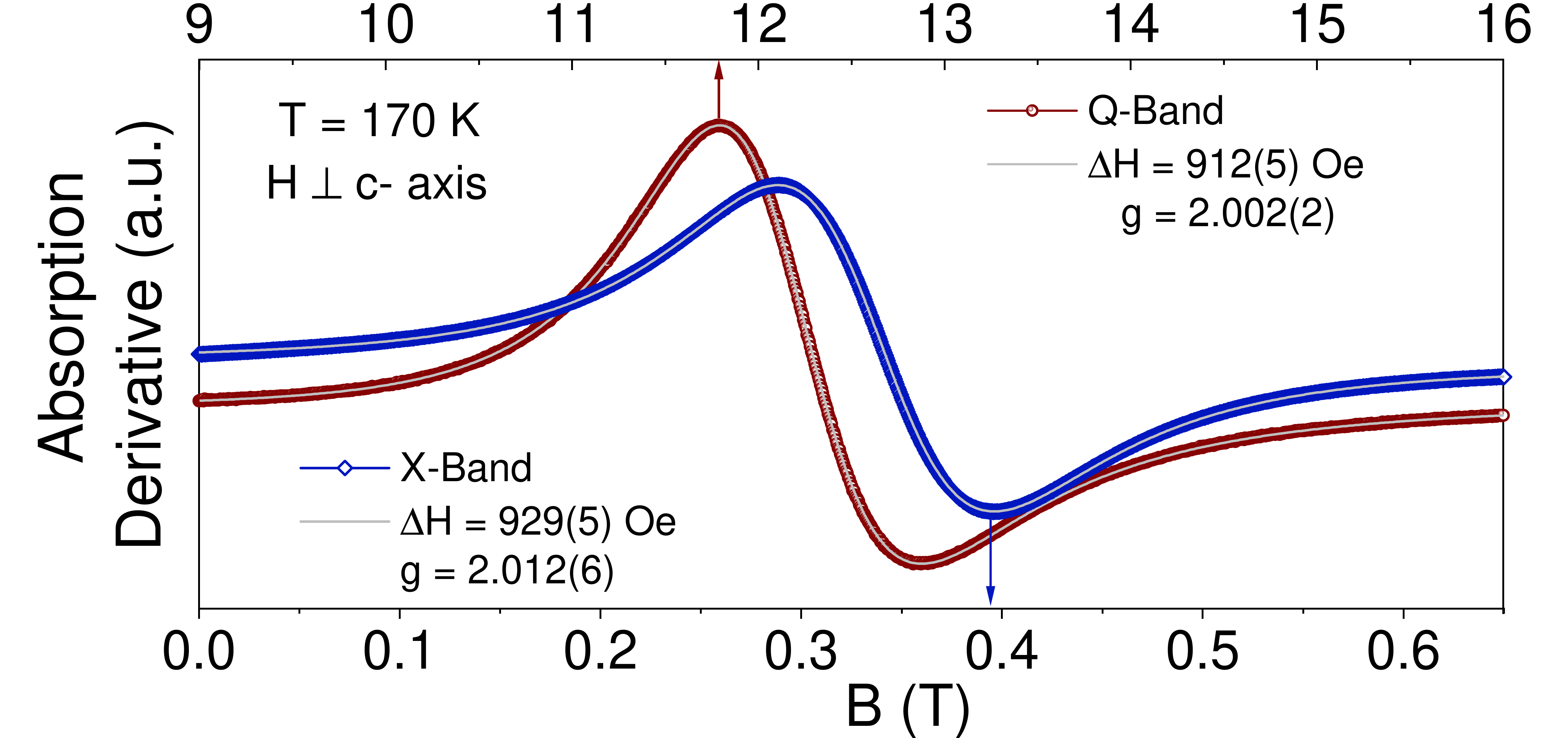}
   \end{center}
   \vspace{-0.75cm}
   \caption{\textbf{Electron spin resonance of Eu$_{5}$In$_{2}$Sb$_{6}$ single crystals.} ESR spectra at 170~K for X ($f = 9.5$~GHz) and Q ($f = 34$~GHz) bands.}
   \label{fig:Fig4}
   \end{figure}
    
\textbf{Electron spin resonance measurements}

We complete our experimental investigation with microscopic electron spin resonance (ESR) measurements. 
ESR is a site-specific spectroscopic technique, and Eu$^{2+}$ ions are
particularly suitable paramagnetic probes because of their $S$-only state \cite{Urbano2004,Rosa2012}.
The Eu$^{2+}$ ESR spectra of Eu$_{5}$In$_{2}$Sb$_{6}$ in the paramagnetic state, shown in Fig.~4, consists of a single unresolved 
resonance (i.e., no fine or hyperfine structure).
The ESR linewidth, $\Delta H$, provides information on the interactions of the spins with their environment and their
motion. In the case of semimetallic EuB$_{6}$, the Eu$^{2+}$ $\Delta H$ was claimed to be 
dominated by spin-flip scattering due to the exchange between 
$4f$ and conduction electrons \cite{Urbano2004}. As a result, $\Delta H$ narrows at higher fields due to a reduction in 
the spin-flip scattering, consistent with the presence of magnetic polarons. The linewidth of Eu$_{5}$In$_{2}$Sb$_{6}$ also 
narrows at higher fields ($Q$-band) when compared to low fields ($X$-band), though not as strongly 
as in EuB$_{6}$ \cite{Urbano2004}. This narrowing further indicates that the resonance
is homogeneous in the paramagnetic state. 
In the case of a small-gap insulator as Eu$_{5}$In$_{2}$Sb$_{6}$, 
the Eu$^{2+}$ ESR linewidth
is dominated by spin-spin interactions \cite{Urbano2004,Huber1998, Yang2004}. 
The resulting relaxation mechanism is set by $T_{2}$, 
the spin-spin relaxation time, which in turn
 is affected by the distribution of Eu-Eu exchange interactions and internal fields. An applied magnetic field 
 causes an increase in $T_{2}$ as the size of the ferromagnetic polaron grows, which results in the observed 
 ESR line narrowing. At the same time, the
 g-value decreases as a function of magnetic field,
 which indicates an antiferromagnetic inter-polaron coupling.
 Therefore, our ESR results are also consistent with the presence of 
 magnetic polarons in
 Eu$_{5}$In$_{2}$Sb$_{6}$.
More detailed ESR measurements will be the focus of a separate study.

\textbf{Band structure calculations}

To shed light on the possible topological nature of the band structure of Eu$_{5}$In$_{2}$Sb$_{6}$, we 
perform band structure calculations in the 
paramagnetic state by taking the $4f$ orbitals of Eu as core
states, as shown in Fig.~5.
Both barium and europium are divalent in the 526 structure, 
and our experimental results imply that europium has a 
well-localized $f$-electron contribution.
One would therefore naively expect that the band structure and topology of Eu$_{5}$In$_{2}$Sb$_{6}$
are similar to that of Ba$_{5}$In$_{2}$Sb$_{6}$, whose topology is not indicated by any symmetry indicators 
but can be characterized by nontrivial connecting pattern in the Wilson bands \cite{Wieder2018}.  

Remarkably, GGA+SOC calculations in the paramagnetic state of 
Eu$_{5}$In$_{2}$Sb$_{6}$
indicate a semimetal state with one extra band inversion
 compared to Ba$_{5}$In$_{2}$Sb$_{6}$ at the $\Gamma$ point.
 Because there are no symmetry-protected band crossings between the valence
and conduction bands at any $k$-point, a $k$-dependent chemical potential can 
be defined, which yields a fully gapped state.
By calculating the topological indices of the bands below the $k$-dependent chemical potential, 
we find that the extra band inversion at $\Gamma$ point yields a strong topological insulator 
with $(z_{2};z_{2w,1}z_{2w2}z_{2w,3})=(1;000)$, where $z_{2}$ is strong index and $z_{2w,i}$ 
is weak index \cite{Fu2007} , as shown in Fig.~5a. Compared with our experimental results, however, 
the {\it ab initio} calculation with the GGA functional incorrectly predicts Eu$_{5}$In$_{2}$Sb$_{6}$ 
to be semimetallic. Considering the possible underestimation of the band gap in semiconductors by the
GGA functional, we have also performed band structure calculations
using the mBJ potential with a coefficient $c_{\mathrm{MBJ}} = 1.18$, which
was obtained self-consistently. As shown in Figure~5b, the band inversion
near the $\Gamma$ point disappears, and a small gap opens along the $\Gamma - Y$ path.
The topological indices $(z_{2};z_{2w,1}z_{2w2}z_{2w,3})$ are computed to be (0;000).
In fact, surface states are not detected by our electrical transport measurements. 
Scanning tunneling microscopy and angle-resolved photoemission measurements will be valuable
to confirm the absence of in-gap states.

 \begin{figure*}[!ht]
   \begin{center}
   \includegraphics[width=2.\columnwidth,keepaspectratio]{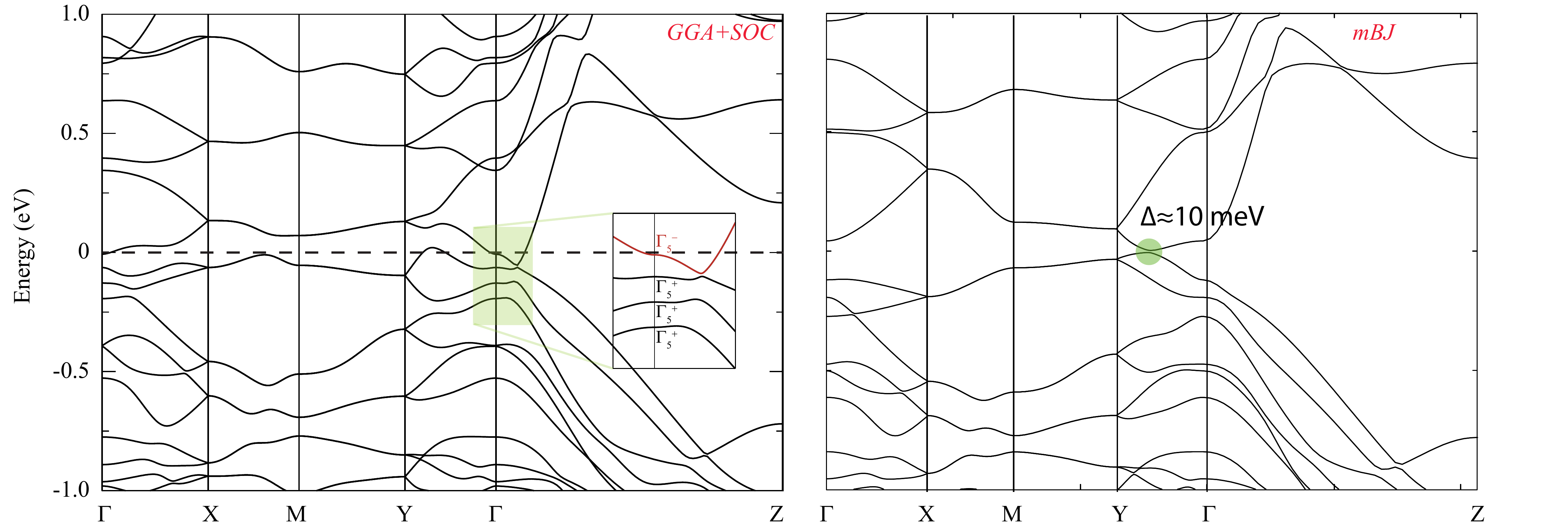}
   \end{center}
   \vspace{-0.25cm}
   \caption{\textbf{Band structure calculations for Eu$_{5}$In$_{2}$Sb$_{6}$.} Calculations in the paramagnetic state
   using the GGA+SOC method are shown in the left panel with the band representations near $\Gamma$ point indicated in the inset. 
   Calculations in the paramagnetic state using the mBJ method with $c_{\mathrm{MBJ}}=1.18$ are shown in the right panel. 
   The band inversion disappears and a 
   trivial band gap of about 10~meV opens along the Y-$\Gamma$ direction.}
   \label{fig:Fig5}
   \end{figure*}

We now investigate the topology of Eu$_{5}$In$_{2}$Sb$_{6}$ in the magnetically ordered state. 
  Because the magnetic structure of Eu$_{5}$In$_{2}$Sb$_{6}$ has not been solved yet,
  we investigate theoretically, using the GGA+U+SOC approach, three A-type AFM phases with the easy axis 
  along different directions. All of the antiferromagnetic phases are theoretically characterized by the so-called
    Type-IV magnetic space groups (MSGs) with inversion symmetry. The magnetic topological quantum 
    chemistry theory therefore describes the topology of these MSGs by an index group 
    ($Z_{4}\times Z_{2}^{3}$), as proposed recently \cite{Yuanfeng2020}. From the calculations detailed in the Supplementary Figure 5, the magnetic moment is about 7 $\mu_{B}/$Eu, and
    the energy difference between the different phases is within 3~meV per unit cell.
     From the results tabulated in Supplementary Table 1, all
    three AFM phases are axion insulators with strong indices 
    $(z_{4},z_{2_{1}},z_{2_{2}},z_{2_{3}})=(2,0,0,0)$. By comparing the band structures for three different
    AFM phases, the polarized $4f$ states do not change the band inversion characteristics of the paramagnetic
    state but induce a small exchange splitting near the Fermi level. Though the AFM structure at low
    temperatures has yet to be determined experimentally, we proposed that this phase is an axion insulator 
candidate that preserves inversion symmetry.

\section{Discussion}

The magnetic polaron picture is fully consistent with our data. At high temperatures ($\sim 15T_{\mathrm{N1}} = 210$~K), the formation of isolated 
magnetic polarons is manifested in magnetic susceptibility measurements via a deviation from the Curie-Weiss law (inset of Fig. ~1a)
and in electrical resistivity
measurements via the onset of negative magnetoresistance (Fig. 2c). 
As the temperature is further lowered, these polarons increase in size until they 
start to interact
 at $T^{*}$ giving rise to a sharp decrease in the
 $\chi T$ plot, 
 a Schottky anomaly in the specific heat data (Fig.~1c), and an anomaly in electrical 
 resistivity measurements (Fig.~2b). 
 At $T_{N1}$, the polarons coalesce and become delocalized, which 
 gives way to a drastic increase in conductivity.
Though the delocalization temperature virtually 
 coincides with $T_{\mathrm{N1}}$ at zero field,
  delocalization is expected to occur at higher temperatures as the size of the polarons increase in field.
  Antiferromagnetic-driven $T^{*}$, however,
 is suppressed in field. This opposite field dependence causes the delocalization temperature and $T^{*}$
 to merge into one at about 3~T, which gives rise to a resistivity maximum above $T_{\mathrm{N1}}$ that moves to
 higher temperatures in field (see Supplementary Figure 7).
 Importantly, the increase in size of magnetic polarons in applied fields also promotes
large negative (termed colossal)
magnetoresistance in the paramagnetic state. In fact, colossal magnetoresistance (CMR)
 sets in at about $200$~K and peaks just above $T_{\mathrm{N1}}$, as shown in Fig.~2c.

 Another characteristic of CMR materials is the scaling of 
 the low-field MR with the square of the reduced magnetization, $\Delta \rho/ \rho_{0}= C(M/M_{sat})^2$, 
 where $M_{sat}$ is the saturation magnetization \cite{Majumdar1998,Furukawa1994}. 
Just above $T_{\mathrm{N1}}$, this scaling is valid and 
yields $C=50$ (inset of Fig.~2c). When electron scattering is 
dominated by magnetic fluctuations, 
the scaling constant $C$ is proportional to $n^{-2/3}$, $n$ being the carrier density \cite{Majumdar1998}.
The scaling constant calculated this way ($n \sim 10^{12}$/cm$^{3}$ at 15~K) 
is four orders of magnitude higher than the experimentally-determined constant, which is an indication of a distinct mechanism. Another notable exception is EuB$_{6}$, 
for which the field-dependent resistivity was argued 
to be dominated by the increase in polaron size with field rather than by the suppression 
of critical scattering \cite{Sullow2000,Sullow2000b}. In fact, recent scanning tunneling microscopy measurements 
have directly imaged the formation of 
magnetic polarons in EuB$_{6}$ \cite{Pohlit2018}. 

In summary, we investigate the thermodynamic and electrical transport properties of single crystalline 
Eu$_{5}$In$_{2}$Sb$_{6}$, a nonsymmorphic Zintl antiferromagnetic insulator. 
Colossal magnetoresistance sets in at temperatures one order of magnitude higher than the magnetic ordering 
temperature, $T_{\mathrm{N1}} = 14$~K, and peaks just above $T_{\mathrm{N1}}$ reaching $-99.7$\% at 3~T and $-99.999$\% at 9~T. 
This is, to our knowledge, the largest CMR observed in a stoichiometric antiferromagnetic compound. 
Our combined electrical transport and microscopic electron spin resonance measurements point to the presence of magnetic polarons that generate an anomalous Hall component.
Our first-principles band structure calculations yield
 an insulating state with trivial topological
  indices in the paramagnetic state, whereas
 an axion insulating state emerges within putative antiferromagnetic states.
Our results highlight that Zintl phases could provide truly
 insulating states in the search for topological insulators, and rare-earth elements
  provide a route for the discovery of topological interacting phenomena. 
  In fact, Zintl Eu$X_{2}$As$_{2}$ ($X=$ In, Sn) have been recently proposed to be antiferromagnetic topological 
  insulators \cite{Xu2019,Li2019}.
  The metallic-like behavior observed in electrical resistivity, however, suggests that these materials have a semimetallic
  ground state akin to EuB$_{6}$ \cite{Nie2019}.

\subsection{Methods}
\textbf{Experimental details}
Single crystalline samples of Eu$_{5}$In$_{2}$Sb$_{6}$ were grown using a combined In-Sb self-flux technique.
The crystallographic
structure was verified at room temperature by both single-crystal diffraction using Mo radiation in a 
commercial diffractometer (see Supplementary Figure 6) and powder diffraction using Cu radiation in a commercial diffractometer. 
Eu$_{5}$In$_{2}$Sb$_{6}$ crystallizes in an orthorhombic structure (space group 55) with
lattice parameters $a = 12.553(5)$\AA, $b = 14.603(2)$\AA$\,$ and $c = 4.635(1)$\AA.
As shown in Supplementary Figure 6, the observed mosaicity of the Bragg reflections
is limited by the resolution of the diffractometer.
The crystals have a rod-like shape, the $c$ axis is the long axis, and typical sizes are 0.5~mm x 0.5mm x 3mm.
In addition, the stoichiometry of crystals was checked by energy dispersive X-ray
spectroscopy (EDX). Magnetization measurements were performed in a commercial SQUID-based
magnetometer. Specific heat measurements were made using the thermal relaxation 
technique in a
commercial measurement system. Because of the difficulties in the synthesis 
of phase pure Ba$_{5}$In$_{2}$Sb$_{6}$, no phonon background was subtracted from the data.
A four-probe configuration was used in the electrical resistivity
experiments performed using a low-frequency AC bridge.
High-field magnetization measurements were performed in the 65~T pulse field magnet at 4~K at the National High magnetic Field Laboratory at Los Alamos National Laboratory. 
Details of the magnetometer design are described in Ref.\cite{Detwiller2000}. 
The sample was mounted in a plastic cup oriented with $b$-axis parallel to the magnetic field. The data were normalizing by the low-field data obtained from a commercial SQUID magnetometer.

ESR measurements were performed on single crystals in X-band ($f = 9.5$ GHz) and Q-band ($f = 34$ GHz) spectrometers
 equipped with a goniometer and a He-flow cryostat in the temperature range of $4~$K $<T < 300$~K.

\textbf{Theoretical details}
First-principle calculations were performed using the Vienna ab initio simulation package (VASP), and 
the generalized gradient approximation (GGA) with 
the Perdew-Burke-Ernzerhof(PBE) type exchange correlation potential was adopted. The Brillouin zone (BZ) 
sampling was performed by using $k$ grids with an $7 \times 7 \times 9$ mesh in self-consistent calculations.
In the paramagnetic state, we employed an europium pseudopotential with seven 
$f$ electrons treated as core electrons. In the antiferromagnetic states, 
we performed the LSDA+U calculations with $U=5$~eV for the three distinct magnetic structures.

 \subsection{Data availability}
 Data presented in this study are available from authors upon request.

\begin{acknowledgments}
We acknowledge constructive discussions with Z. Fisk, S. Wirth, J. Muller, O. Erten and C. Kurdak.
Synthesis and macroscopic measurements at low fields were supported by the U.S. Department of Energy (DOE)  
BES ``Quantum Fluctuations in Narrow-Band Systems" project.
High-field magnetization measurements were supported by the DOE BES ``Science of 100 Tesla" project. 
SKK acknowledges support from the LANL Director's Postdoctoral LDRD program. The National High Magnetic Field Laboratory 
is supported by National Science Foundation through NSF/DMR-1644779 and the State of Florida.
Scanning electron microscope and energy dispersive X-ray measurements were performed at the Center 
for Integrated Nanotechnologies, an Office of Science User Facility operated for the DOE
Office of Science. Electron spin resonance measurements were supported by FAPESP (SP-Brazil) grants no. 2018/11364-7, 2017/10581-1, 2012/04870-7, 
CNPq Grant no. 141026/2017-0, CAPES and FINEP-Brazil.
We thank Diamond Light Source for the provision of beamtime under proposal MT18991.
Work at UCL is supported by the UK Engineering and Physical Sciences Research Council (Grants No. EP/N027671/1 and No. EP/N034694/1)
Theory efforts were supported by DOE de-sc0016239, NSF EAGER 1004957, Simons Investigator Grants, 
ARO MURI W911NF- 12-1-0461, the Packard Foundation, and the Schmidt Fund for Innovative Research.
Y.X. and B.A.B. were supported by Max Planck society.
Z. W. received support from the National Natural Science Foundation of China (No. 11974395), 
the Chinese Academy of Sciences (CAS) [Grant No. XDB33000000], and the CAS Pioneer Hundred Talents Program.
\end{acknowledgments}

\subsection{Author Contributions}

PFSR and FR conceived the project. PFSR and EDB synthesized single crystals for the measurement. 
 PFSR and SMT performed electrical transport and specific heat measurements. JDT performed magnetization measurements to 6~T.
 SKK, MKC, and NH performed magnetization measurements to 60~T. 
 JCZ and PGP performed electron spin resonance measurements.
 MCR, LSIV, MJ, and AB performed x-ray absorption measurements measurements. 
YX, ZW, and BAB performed band structure calculations.
PFSR wrote the manuscript with input from all authors.

 \subsection{Competing Interests}
 The Authors declare no competing financial or non-financial interests.

\bibliographystyle{naturemag}

\end{document}